\documentclass[superscriptaddress,reprint,amsmath,amssymb,aps,floatfix]{revtex4-1}

\usepackage{amsmath,gensymb,textcomp,bm,dcolumn,eurosym,array,tabu,multirow,nicefrac,color,subfigure,graphicx,upgreek}
\usepackage[colorlinks, linkcolor=blue, citecolor=blue, urlcolor=blue, breaklinks=true]{hyperref}

\usepackage{mathpazo,times} 

\newcommand{\ket}[1]{\lvert #1 \rangle}

\usepackage[subsectionbib]{bibunits}
\usepackage{graphicx}
\usepackage{dcolumn}
\usepackage{bm,MnSymbol}
\usepackage{units}
\usepackage{psfrag}
\usepackage{float}
\usepackage{color}

\begin{document}


\title{Polarization entanglement by time-reversed Hong-Ou-Mandel interference}




\author{Yuanyuan Chen}
\email{chenyy@smail.nju.edu.cn}
\affiliation{
Institute for Quantum Optics and Quantum Information - Vienna (IQOQI), Austrian Academy of Sciences, Boltzmanngasse 3, 1090 Vienna, Austria.}
\affiliation{Vienna Center for Quantum Science and Technology (VCQ), Vienna, Austria}
\affiliation{State Key Laboratory for Novel Software Technology, Nanjing University,
Xianlin Avenue 163, Nanjing 210046, China.}

\author{Sebastian Ecker}
\affiliation{
Institute for Quantum Optics and Quantum Information - Vienna (IQOQI), Austrian Academy of Sciences, Boltzmanngasse 3, 1090 Vienna, Austria.}
\affiliation{Vienna Center for Quantum Science and Technology (VCQ), Vienna, Austria}

\author{S\"oren Wengerowsky}
\affiliation{
Institute for Quantum Optics and Quantum Information - Vienna (IQOQI), Austrian Academy of Sciences, Boltzmanngasse 3, 1090 Vienna, Austria.}
\affiliation{Vienna Center for Quantum Science and Technology (VCQ), Vienna, Austria}

\author{Lukas Bulla}
\affiliation{
Institute for Quantum Optics and Quantum Information - Vienna (IQOQI), Austrian Academy of Sciences, Boltzmanngasse 3, 1090 Vienna, Austria.}
\affiliation{Vienna Center for Quantum Science and Technology (VCQ), Vienna, Austria}

\author{Siddarth Koduru Joshi}
\thanks{Current address: Quantum Engineering Technology Labs, H. H. Wills Physics Laboratory \& Department of Electrical and
Electronic Engineering, University of Bristol, Merchant Venturers Building, Woodland Road, Bristol BS8 1UB, United Kingdom}
\affiliation{
Institute for Quantum Optics and Quantum Information - Vienna (IQOQI), Austrian Academy of Sciences, Boltzmanngasse 3, 1090 Vienna, Austria.}
\affiliation{Vienna Center for Quantum Science and Technology (VCQ), Vienna, Austria}

\author{Fabian Steinlechner}
\affiliation{
Institute for Quantum Optics and Quantum Information - Vienna (IQOQI), Austrian Academy of Sciences, Boltzmanngasse 3, 1090 Vienna, Austria.}
\affiliation{Vienna Center for Quantum Science and Technology (VCQ), Vienna, Austria}

\author{Rupert Ursin}
\email{Rupert.Ursin@oeaw.ac.at}
\affiliation{
Institute for Quantum Optics and Quantum Information - Vienna (IQOQI), Austrian Academy of Sciences, Boltzmanngasse 3, 1090 Vienna, Austria.}
\affiliation{Vienna Center for Quantum Science and Technology (VCQ), Vienna, Austria}


%


\begin{abstract} 
Sources of entanglement are an enabling resource in quantum technology, and pushing the limits of generation rate and quality of entanglement is a necessary pre-requisite towards practical applications. Here, we present an ultra-bright source of polarization-entangled photon pairs based on time-reversed Hong-Ou-Mandel interference. By superimposing four pair-creation possibilities on a polarization beam splitter, pairs of identical photons are separated into two spatial modes without the usual requirement for wavelength distinguishability or non-collinear emission angles.
Our source yields high-fidelity polarization entanglement and high pair-generation rates without any requirement for active interferometric stabilization, which makes it an ideal candidate for a variety of applications, in particular those requiring indistinguishable photons. 
\end{abstract}
\maketitle

\indent \emph{Introduction.}\rule[2pt]{8pt}{1pt}Quantum entanglement is an enabling resource for quantum information processing (QIP) and an efficient source of entangled photons can now be considered an absolute necessity in the quantum mechanic's toolkit. Entangled photons can be generated using a variety of technological approaches \cite{de2016quantum}, with spontaneous parametric down-conversion (SPDC) in nonlinear materials representing the present-day gold standard with respect to fiber coupling efficiency \cite{giustina2015significant,shalm2015strong}, entangled photon pair rates \cite{steinlechner2013phase}, and entanglement fidelity \cite{poh2015approaching}. In the SPDC process, photons from a strong pump laser (p) spontaneously decay into two daughter photons, commonly referred to as signal (s) and idler (i), which can be tailored to exhibit entanglement in various photonic degrees of freedom.
SPDC offers a wide range of possibilities to generate polarization entanglement \cite{kwiat1995new,kwiat1999ultrabright,shi2004generation,fiorentino2005source,kim2006phase,lee2016polarization,villar2018experimental}, with two widely used source configurations being the crossed-crystal scheme, in which two parametric down-converters, rotated by $90^{\circ}$  with respect to each other, are placed in sequence and pumped with a diagonally polarized pump laser \cite{kwiat1999ultrabright}, and the Sagnac scheme, where a single down-converter is bi-directionally pumped inside a polarization Sagnac interferometer (PSI) \cite{shi2004generation,kim2006phase}. These schemes owe their popularity to the fact that no active interferometric stabilization is required, due to the common path configuration for down-converted and pump photons.\\
\indent The past two decades have seen significant efforts dedicated to improving the efficiency and tunability of SPDC sources. In particular the advent of periodic poling technology has greatly extended the range of possible SPDC configurations, and made it possible to engineer highly efficient collinear quasi-phase matching (QPM) in long periodically poled nonlinear crystals \cite{kim2006phase,fedrizzi2007wavelength,steinlechner2012high,steinlechner2013phase,steinlechner2014efficient,lee2016polarization,jabir2017robust,dietz2016folded} and waveguide structures \cite{fiorentino2007spontaneous,sohler2012integrated,krapick2013efficient,vergyris2017fully,matsuda2012monolithically,clausen2014source}. In particular, SPDC sources that exploit the strong nonlinear interaction of collinear QPM with co-polarized pump, signal, and idler photons, so-called type-0 QPM, have resulted in the highest entangled pair rates reported to date \cite{steinlechner2015sources,steinlechner2013phase}. Due to the spatial overlap of the SPDC and pump modes, however, these type-0 sources typically require wavelength distinguishability ($\lambda_s\neq \lambda_i$) or non-collinear SPDC emission angles in order to route photons into distinct spatial modes for independent manipulation (e.g. using dichroic mirrors or fiber-Bragg gratings). As this limits their applicability in QIP protocols that require indistinguishable photons, the question naturally arises, how we may generate polarization-entanglement such that identical signal and idler photons are separated deterministically. That is, conditional on the detection of a signal (idler) photon in one spatial mode, there should be, in principle, a unit probability of a corresponding detection of the idler (signal) photon in a conjugate spatial mode. In contrast to this ideal case, the widely used probabilistic separation on a beam splitter always results in an undesirable two-photon component in its output ports.\\
\indent To tackle this issue, Chen et al. \cite{chen2007deterministic} proposed a deterministic \emph{quantum splitter} that uses two-photon interference to passively route photons into two spatial modes. The most widely known manifestation of two-photon interference is the Hong-Ou-Mandel (HOM) effect, where identical photons impinging on the input ports of a beam splitter bunch into either one or the other output port. In a time-reversed analogy, interference of two indistinguishable photon pairs results in anti-bunching in the output ports of the beam splitter, i.e. a suppression of two-photon components in each port. This quantum splitter approach has since found applications in several experiments, in particular on integrated waveguide platforms \cite{jin2014chip,marchildon2016deterministic}.\\
\indent Here we present a novel source configuration which exploits time-reversed HOM interference to deterministically route wavelength-degenerate polarization-entangled photons into two distinct spatial modes. By coherent superposition of identical photon pairs in four different modes on a polarization beam splitter, our source yields polarization entanglement without any requirement for detection post-selection.  This is achieved in a phase-stable manner by combining the benefits of the popular polarization Sagnac sources and crossed-crystal sources. Using highly efficient collinear type-0 QPM in bulk periodically poled potassium titanyl phosphate (ppKTP) crystals, we generate wavelength-degenerate photon pairs around $\unit[810]{nm}$ with a Bell-state fidelity of $99.2\%$ and detect a pair rate of $\unit[160]{kcps}$ per mW of pump power.\\
\indent This photon pair yield, which -- to the best of our knowledge -- is the highest value reported for a wavelength-degenerate polarization-entangled photon source, is of particular relevance whenever the available pump power is limited, as is the case in space-proof entangled photon sources for satellite-based quantum communication \cite{yin2017satellite,liao2017satellite,liao2017long}, or fundamental tests of quantum theory in scenarios with extreme link loss  \cite{scheidl2013quantum,wang2014link,neiman2018towards}. Moreover, the scheme can be extended to integrated source platforms, where the separation of photons generated in overlapping, co-propagating spatial modes is particularly challenging.

\begin{figure}[!ht]
\centering
\includegraphics[width=\linewidth]{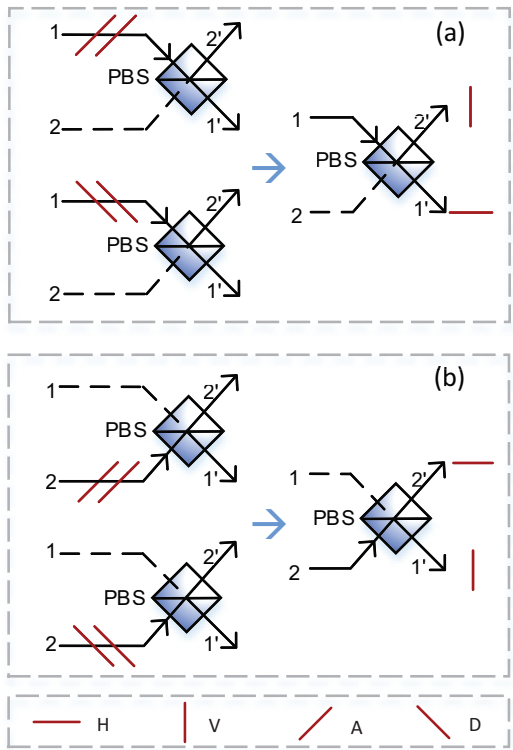}
\caption{Illustration of the source's
principle. A pair of identical photons in a correlated state is incident on a PBS via either input port 1 (Fig. 1(a)) or 2 (Fig. 1(b)). As a consequence of time-reversed HOM interference of orthogonal polarization modes, the photons anti-bunch into the output ports 1' and 2'. Superposition of (a) and (b) thus results in a polarization-entangled state. H, V, A and D represent horizontal, vertical, anti-diagonal and diagonal polarizations, respectively.}
\label{figure_1}
\end{figure}

\indent \emph{Basic scheme.}\rule[2pt]{8pt}{1pt}To illustrate the operational principle of the source (Fig. \ref{figure_1}), let us consider a pair of identical photons which are either both linearly polarized along the diagonal or anti-diagonal direction. The photons are incident on a polarization beam splitter (PBS) in a state:

\begin{equation}\label{eq:crossed-crystal}
\frac{({a_D^{\dagger}}^{2}+ e^{i\phi} {a_A^{\dagger}}^{2})}{\sqrt{2}} \ket{\text{vac}}_{1} = \frac{\ket{2_D,0_A}_{1} + e^{i\phi}  \ket{0_D,2_A}_{1}}{\sqrt{2}},
\end{equation}

where $a^{\dagger}_D=1 / \sqrt{2}(a^{\dagger}_H+a^{\dagger}_V)$ and $a^{\dagger}_A=1/\sqrt{2}(a^{\dagger}_H-a^{\dagger}_V)$ represent the creation operators for photons polarized diagonally and anti-diagonally with respect to the rectilinear reference frame of the PBS (Fig. 1(a)). The vacuum state is represented by $\ket{\text{vac}}$ and $\phi$ represents a polarization-dependent phase factor. Setting the relative phase $\phi=\pi$, the state reads:
 
\begin{equation}\label{eq:HOM}
\frac{\ket{2_D,0_A}_{1}-\ket{0_D,2_A}_{1}}{\sqrt{2}} =\ket{1_H,1_V}_{1}.
\end{equation}

This can be seen as anti-bunching due to Hong-Ou-Mandel interference in the orthogonal polarization modes \cite{ray2011verifying,buse2015photons}. The  PBS then maps the orthogonally polarized photon pairs into two distinct spatial modes: 

\begin{equation}
\ket{1_H,1_V}_{1}\rightarrow \ket{H}_{1'}\ket{V}_{2'} \equiv \ket{\Psi^{(1)}}, 
\end{equation}

where $\ket{H}_{i}\equiv	
\ket{1_H,0_V}_{i}$ and $\ket{V}_{i}\equiv	
\ket{0_H,1_V}_{i}$ denote horizontal and vertical single-photon polarization states in ports $i=\{1',2'\}$, respectively.

Analogously, when two photons in a state (\ref{eq:HOM}) are incident via port 2 of the PBS (Fig. 1(b)) one obtains single photons with the orthogonal polarization state:

\begin{equation}
\ket{1_H,1_V}_{2} \rightarrow \ket{V}_{1'}\ket{H}_{2'} \equiv \ket{\Psi^{(2)}},
\end{equation}

Consequently, the coherent superposition of pair generation possibilities $\ket{\Psi^{(1)}}$ + $\ket{\Psi^{(2)}}$ results in a maximally polarization-entangled state in spatial modes 1' and 2':

\begin{equation}\label{eq:psi}
\ket{\Psi}_{1',2'} = \frac{1}{\sqrt{2}}\left(\ket{H_{1'}V_{2'}}+\ket{V_{1'}H_{2'}}\right).
\end{equation}

In our realization of this operational principle we  produce pairs of photons in a state \eqref{eq:crossed-crystal} by balanced pumping a pair of crossed crystals with a relative inclination about their common propagation axis of $90^\circ$. By folding the configuration depicted in Fig. \ref{figure_1} into a loop we realize spatial modes 1 and 2 as the clockwise- and counter-clockwise propagation modes of a PSI. The polarization-entangled state \eqref{eq:psi} is then obtained by bi-directionally pumping the two crystals, which are placed in the center of the loop. 
This implementation thus ensures constant phases in states \eqref{eq:crossed-crystal} and \eqref{eq:psi} without any requirement for active interferometric stabilization (changes of the optical path length are experienced by the pump driving the SPDC process as well as the emitted biphoton state). Another benefit reveals itself when considering the multi-mode spatio-temporal characteristics of the bi-photon wavepackets \cite{trojek2008collinear}; due to the symmetric Sagnac configuration, there is, in principle, no requirement to remove distinguishing arrival-time information (e.g. via birefringent compensation crystals) as it is never created in the first place. Hence, the scheme can, in principle, also be extended to large SPDC bandwidths, and even non-degenerate wavelengths. \\
\indent Note that, while the scheme can also be realized using a non-polarizing beam splitter, the implementation with a PBS automatically ensures that the correct polarizations are sorted into the two output ports, thus improving the fidelity of the polarization-entangled state. This feature can also be interpreted as an entanglement purification step \cite{pan2003experimental}, wherein the state impurity due to residual distinguishing information in the interfering modes merely affects the anti-bunching probability. Imperfect indistinguishability thus reduces the anti-bunching probability, and consequently the rate of joint detections in spatial modes 1' and 2'. However, it should in principle not affect the quality of the polarization correlations conditioned on a joint detection in these two modes.


\begin{figure}[!t]
\centering
\includegraphics[width=\linewidth]{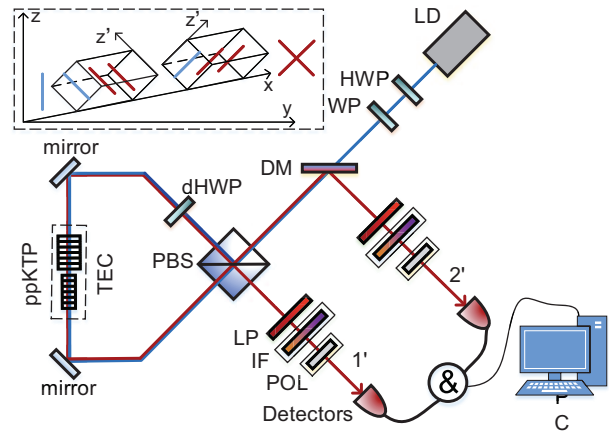}
\caption{Illustration of the crossed-crystal Sagnac source. LD: laser diode; PBS: polarization beam splitter; dHWP: dual-wavelength half wave plate; HWP: half wave plate; WP: wave plate; DM: dichroic mirror; ppKTP: type-0 periodically poled potassium titanyl phosphate crystal; TEC: temperature controller; LP: long pass filter; IF: interference filter; POL: polarizer. The top left inset illustrates the $45^\circ$ orientation of the oven which ensures that photon pairs are generated either with diagonal or anti-diagonal polarizations.}
\label{figure_2}
\end{figure}

\indent \emph{Experiment.}\rule[2pt]{8pt}{1pt}In our experimental realization of the source design (Fig. \ref{figure_2}) a pair of crossed ppKTP crystals is placed inside a Sagnac loop configuration and pumped with a $\unit[405]{nm}$ continuous wave grating-stabilized laser diode. A half-wave plate (HWP) in the pump beam is used to set a diagonal polarization state, such that both clockwise and counter-clockwise directions of the interferometer are pumped equally. To achieve the desired diagonal and anti-diagonal polarizations within the Sagnac loop, we designed an oven with a V-groove such that the two crossed crystals are oriented at $\unit[45]{^\circ}$ (see inset of Fig. \ref{figure_2}). Thus, the crystals are phase-matched for SPDC with diagonally and anti-diagonally polarized pump light, respectively. The crossed-crystal configuration at the center of the loop is based on two mutually orthogonally oriented 11.48-mm-long ppKTP crystals. They provide type-0 collinear phase matching with pump (p), signal (s) and idler (i) photons at center wavelengths of $\lambda_{p}\approx \unit[405]{nm}$ and $\lambda_{s,i}\approx \unit[810]{nm}$ at a temperature of $\unit[107]{^\circ C}$. Since the pump beam in the clockwise/counter-clockwise propagation direction is horizontally/vertically polarized, it is equally likely to generate a photon pair in the first crystal or the second crystal, resulting in a state of \eqref{eq:crossed-crystal}. The relative polarization phase was tuned by tilting a wave-plate (WP) with optical axis set at $45^{\circ}$. 


After combining the SPDC photon pairs from both propagation directions on the PBS, the down converted signal and idler photons in spatial mode 2' are separated from the pump by using a dichroic mirror and coupled into single mode fibers. Two long-pass filters are used to eliminate the remaining pump light and noise. Polarizers are used to evaluate  polarization correlations and bandpass filters are utilized to adjust the spectral bandwidth of the generated entangled state. Finally, the down converted photons are detected by silicon avalanche photo diodes, and two-fold events are identified using a fast electronic AND gate when two photons arrive at the detectors within a coincidence window of $\sim \unit[3]{ns}$.

\indent \emph{Results.}\rule[2pt]{8pt}{1pt}In order to evaluate the source brightness, we removed the polarizers from the setup and set the pump laser power to approximately $\unit[100]{\micro W}$. With two $\unit[3]{nm}$ bandpass filters in place, we detect a two-fold coincidence rate of $R_{c}\approx 16$ kcps and single count rates of $R_{s}\approx R_{i}\approx 86$ kcps. This corresponds to a normalized pair rate of 160 kcps/mW, a spectral brightness of 53 kcps/mW/nm and a heralding efficiency of  $\frac{R_{c}}{R_{s}}\approx\frac{R_{c}}{R_{i}}\sim18.5\%$ for the idler and signal photons. 
 
Next, we characterized the polarization entanglement by measuring the two-photon polarization interference contrast in two mutually unbiased bases. We observe fringe visibilities of $V_{H/V}=99.3\%(99.7\%)$ in the H/V basis and $V_{A/D}=98.1\%(98.6\%)$ in the A/D basis without (with) subtraction of accidental coincidences. 
These visibilities imply lower bounds of $F\geq  0.992$ and $\mathcal{C}\geq 0.984$ on the Bell-state Fidelity and  Concurrence, respectively \cite{steinlechner2017distribution}.
 
We also assessed the source performance without the additional $\unit[3]{nm}$ bandpass filters in place. Collecting photon pairs for the entire phase-matching bandwidth for wavelength-degenerate type-0 SPDC ($\sim \unit[20]{nm}$ FWHM), the normalized pair rate is increased to 1.07 Mcps/mW and the spectral brightness remains essentially unchanged at $53$ kcps/mW/nm.  However, the Bell-state fidelity is reduced to F $\sim$ 0.88. This reduction is mainly due to the diminished visibility in the A/D basis of $78\%$, which indicates that there remains a residual degree of  distinguishability in the time-frequency domain.

In a perfectly symmetric Sagnac loop, the relative phase of clockwise and counter-clockwise propagating pairs should be perfectly matched. We thus attribute the remaining wavelength dependent phase to non-ideal optical components; the most likely candidates being either polarization-dependent group velocity dispersion of the broad-band multi-layer mirror coatings, or the dual-wavelength PBS, which was designed for a narrow wavelength range around $\unit[810]{nm}$. We believe that it should be possible to obtain high visibility for the full spectrum by incorporating appropriate zero-phase-shift optical components.
  
\begin{figure}[!t]
\centering
\includegraphics[width=\linewidth]{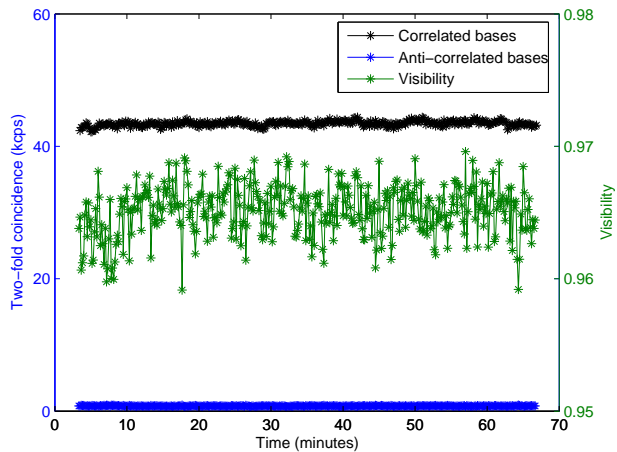}
\caption{Stability of our source over long time under laboratory conditions. The two-fold coincidence counts and visibilities in correlated and anti-correlated A/D bases are stable on the order of hours, indicating its suitability for long-term operation in field experiments.}
\label{figure_3}
\end{figure}

\indent To verify the long-time stability of our source, we performed measurements in the correlated and anti-correlated A/D polarization bases over the course of an hour (Fig. \ref{figure_3}). The results illustrate the good temporal stability of the source efficiency, as well as the quality of the entangled state, making it a suitable source for long-term autonomous operation. 

\indent \emph{Discussion.}\rule[2pt]{8pt}{1pt}We have demonstrated a novel entangled photon source configuration based on time-reversed HOM interference to passively route indistinguishable photons into two spatial modes. The proposed source allows the use of efficient type-0 SPDC in bulk ppKTP in a wavelength-degenerate QPM scheme, without the usual requirement for detection post-selection. In particular, the recent advances in satellite-based quantum communication \cite{yin2017satellite,liao2017satellite,liao2017long}, and proposals for space link experiments with extreme loss \cite{scheidl2013quantum,wang2014link,neiman2018towards} have highlighted the need for ultra-bright, resource-efficient quantum sources with compact footprint and long-term stability.

Our source yields entangled photon rates in excess of $10^7$ pairs per second for pump powers readily attainable using compact laser diodes, making it an ideal candidate for a variety of applications. We note that, using a more tightly focused pump beam we could drastically improve the brightness; in principle, we expect our source to be capable of producing pair rates as high as those reported for non-degenerate phase matching \cite{steinlechner2012high,steinlechner2013phase}, which could be of particular relevance in the development of ultra-bright space-proof entangled photon sources. 

While our experimental realization is based on bulk optics, the overall scheme could also be extended to integrated quantum sources, where the separation of co-propagating photons with overlapping spatial modes poses a significant challenge.

We also note that the time-reversed HOM could also be extended to yield other forms of entanglement, e.g. frequency-polarization hyperentanglement or high-dimensional orbital angular momentum entanglement \cite{zhang2016engineering}. Another promising line of inquiry could address the combination with entanglement by path identity  \cite{krenn2017entanglement}. Finally, we also envisage the use of this approach in engineering multi-photon entanglement \cite{scheidl2014crossed}, where unbalanced beam splitting ratios could enable the filtration and tailoring of desired photon number characteristics \cite{sanaka2006filtering,resch2007entanglement}. 

In conclusion, we hope that our results will inspire new experimental configurations based on multi-photon interference. We believe that fully harnessing HOM interference could provide valuable generating and detecting complex forms of high-dimensional multi-photon entanglement, as required for the next generation of multi-partite quantum information processing protocols.

\vspace{0.5cm}
\indent We thank Thomas Scheidl and Johannes Handsteiner for helpful conversations and comments on the initial draft of the manuscript as well as Valerio Pruneri for providing ppKTP crystals. YC thanks Lijun Chen for support.
Financial support from the Austrian Research Promotion Agency (FFG) Projects - Agentur f\"{u}r Luft- und Raumfahrt (FFG-ALR contract 6238191 and 866025), the European Space Agency (ESA contract 4000112591/14/NL/US) as well as the Austrian Academy of Sciences is gratefully acknowledged. YC acknowledges personal funding from Major Program of National Natural Science Foundation of China (No. 11690030, 11690032), National Key Research and Development Program of China (2017YFA0303700); the National Natural Science Foundation of China (No.61771236), and from a Scholarship from the China Scholarship Council
(CSC).

\bibliography{apssamp}

\providecommand{\noopsort}[1]{}\providecommand{\singleletter}[1]{#1}%
\begin{thebibliography}{43}%
\makeatletter
\providecommand \@ifxundefined [1]{%
 \@ifx{#1\undefined}
}%
\providecommand \@ifnum [1]{%
 \ifnum #1\expandafter \@firstoftwo
 \else \expandafter \@secondoftwo
 \fi
}%
\providecommand \@ifx [1]{%
 \ifx #1\expandafter \@firstoftwo
 \else \expandafter \@secondoftwo
 \fi
}%
\providecommand \natexlab [1]{#1}%
\providecommand \enquote  [1]{``#1''}%
\providecommand \bibnamefont  [1]{#1}%
\providecommand \bibfnamefont [1]{#1}%
\providecommand \citenamefont [1]{#1}%
\providecommand \href@noop [0]{\@secondoftwo}%
\providecommand \href [0]{\begingroup \@sanitize@url \@href}%
\providecommand \@href[1]{\@@startlink{#1}\@@href}%
\providecommand \@@href[1]{\endgroup#1\@@endlink}%
\providecommand \@sanitize@url [0]{\catcode `\\12\catcode `\$12\catcode
  `\&12\catcode `\#12\catcode `\^12\catcode `\_12\catcode `\%12\relax}%
\providecommand \@@startlink[1]{}%
\providecommand \@@endlink[0]{}%
\providecommand \url  [0]{\begingroup\@sanitize@url \@url }%
\providecommand \@url [1]{\endgroup\@href {#1}{\urlprefix }}%
\providecommand \urlprefix  [0]{URL }%
\providecommand \Eprint [0]{\href }%
\providecommand \doibase [0]{http://dx.doi.org/}%
\providecommand \selectlanguage [0]{\@gobble}%
\providecommand \bibinfo  [0]{\@secondoftwo}%
\providecommand \bibfield  [0]{\@secondoftwo}%
\providecommand \translation [1]{[#1]}%
\providecommand \BibitemOpen [0]{}%
\providecommand \bibitemStop [0]{}%
\providecommand \bibitemNoStop [0]{.\EOS\space}%
\providecommand \EOS [0]{\spacefactor3000\relax}%
\providecommand \BibitemShut  [1]{\csname bibitem#1\endcsname}%
\let\auto@bib@innerbib\@empty
\bibitem [{\citenamefont {De~Touzalin}\ \emph {et~al.}(2016)\citenamefont
  {De~Touzalin}, \citenamefont {Marcus}, \citenamefont {Heijman}, \citenamefont
  {Cirac}, \citenamefont {Murray},\ and\ \citenamefont
  {Calarco}}]{de2016quantum}%
  \BibitemOpen
  \bibfield  {author} {\bibinfo {author} {\bibfnamefont {A.}~\bibnamefont
  {De~Touzalin}}, \bibinfo {author} {\bibfnamefont {C.}~\bibnamefont {Marcus}},
  \bibinfo {author} {\bibfnamefont {F.}~\bibnamefont {Heijman}}, \bibinfo
  {author} {\bibfnamefont {I.}~\bibnamefont {Cirac}}, \bibinfo {author}
  {\bibfnamefont {R.}~\bibnamefont {Murray}}, \ and\ \bibinfo {author}
  {\bibfnamefont {T.}~\bibnamefont {Calarco}},\ }\href@noop {} {\bibfield
  {journal} {\bibinfo  {journal} {European Comission}\ } (\bibinfo {year}
  {2016})}\BibitemShut {NoStop}%
\bibitem [{\citenamefont {Giustina}\ \emph {et~al.}(2015)\citenamefont
  {Giustina}, \citenamefont {Versteegh}, \citenamefont {Wengerowsky},
  \citenamefont {Handsteiner}, \citenamefont {Hochrainer}, \citenamefont
  {Phelan}, \citenamefont {Steinlechner}, \citenamefont {Kofler}, \citenamefont
  {Larsson}, \citenamefont {Abell{\'a}n} \emph
  {et~al.}}]{giustina2015significant}%
  \BibitemOpen
  \bibfield  {author} {\bibinfo {author} {\bibfnamefont {M.}~\bibnamefont
  {Giustina}}, \bibinfo {author} {\bibfnamefont {M.~A.}\ \bibnamefont
  {Versteegh}}, \bibinfo {author} {\bibfnamefont {S.}~\bibnamefont
  {Wengerowsky}}, \bibinfo {author} {\bibfnamefont {J.}~\bibnamefont
  {Handsteiner}}, \bibinfo {author} {\bibfnamefont {A.}~\bibnamefont
  {Hochrainer}}, \bibinfo {author} {\bibfnamefont {K.}~\bibnamefont {Phelan}},
  \bibinfo {author} {\bibfnamefont {F.}~\bibnamefont {Steinlechner}}, \bibinfo
  {author} {\bibfnamefont {J.}~\bibnamefont {Kofler}}, \bibinfo {author}
  {\bibfnamefont {J.-{\AA}.}\ \bibnamefont {Larsson}}, \bibinfo {author}
  {\bibfnamefont {C.}~\bibnamefont {Abell{\'a}n}},  \emph {et~al.},\
  }\href@noop {} {\bibfield  {journal} {\bibinfo  {journal} {Physical review
  letters}\ }\textbf {\bibinfo {volume} {115}},\ \bibinfo {pages} {250401}
  (\bibinfo {year} {2015})}\BibitemShut {NoStop}%
\bibitem [{\citenamefont {Shalm}\ \emph {et~al.}(2015)\citenamefont {Shalm},
  \citenamefont {Meyer-Scott}, \citenamefont {Christensen}, \citenamefont
  {Bierhorst}, \citenamefont {Wayne}, \citenamefont {Stevens}, \citenamefont
  {Gerrits}, \citenamefont {Glancy}, \citenamefont {Hamel}, \citenamefont
  {Allman} \emph {et~al.}}]{shalm2015strong}%
  \BibitemOpen
  \bibfield  {author} {\bibinfo {author} {\bibfnamefont {L.~K.}\ \bibnamefont
  {Shalm}}, \bibinfo {author} {\bibfnamefont {E.}~\bibnamefont {Meyer-Scott}},
  \bibinfo {author} {\bibfnamefont {B.~G.}\ \bibnamefont {Christensen}},
  \bibinfo {author} {\bibfnamefont {P.}~\bibnamefont {Bierhorst}}, \bibinfo
  {author} {\bibfnamefont {M.~A.}\ \bibnamefont {Wayne}}, \bibinfo {author}
  {\bibfnamefont {M.~J.}\ \bibnamefont {Stevens}}, \bibinfo {author}
  {\bibfnamefont {T.}~\bibnamefont {Gerrits}}, \bibinfo {author} {\bibfnamefont
  {S.}~\bibnamefont {Glancy}}, \bibinfo {author} {\bibfnamefont {D.~R.}\
  \bibnamefont {Hamel}}, \bibinfo {author} {\bibfnamefont {M.~S.}\ \bibnamefont
  {Allman}},  \emph {et~al.},\ }\href@noop {} {\bibfield  {journal} {\bibinfo
  {journal} {Physical review letters}\ }\textbf {\bibinfo {volume} {115}},\
  \bibinfo {pages} {250402} (\bibinfo {year} {2015})}\BibitemShut {NoStop}%
\bibitem [{\citenamefont {Steinlechner}\ \emph {et~al.}(2013)\citenamefont
  {Steinlechner}, \citenamefont {Ramelow}, \citenamefont {Jofre}, \citenamefont
  {Gilaberte}, \citenamefont {Jennewein}, \citenamefont {Torres}, \citenamefont
  {Mitchell},\ and\ \citenamefont {Pruneri}}]{steinlechner2013phase}%
  \BibitemOpen
  \bibfield  {author} {\bibinfo {author} {\bibfnamefont {F.}~\bibnamefont
  {Steinlechner}}, \bibinfo {author} {\bibfnamefont {S.}~\bibnamefont
  {Ramelow}}, \bibinfo {author} {\bibfnamefont {M.}~\bibnamefont {Jofre}},
  \bibinfo {author} {\bibfnamefont {M.}~\bibnamefont {Gilaberte}}, \bibinfo
  {author} {\bibfnamefont {T.}~\bibnamefont {Jennewein}}, \bibinfo {author}
  {\bibfnamefont {J.~P.}\ \bibnamefont {Torres}}, \bibinfo {author}
  {\bibfnamefont {M.~W.}\ \bibnamefont {Mitchell}}, \ and\ \bibinfo {author}
  {\bibfnamefont {V.}~\bibnamefont {Pruneri}},\ }\href@noop {} {\bibfield
  {journal} {\bibinfo  {journal} {Optics express}\ }\textbf {\bibinfo {volume}
  {21}},\ \bibinfo {pages} {11943} (\bibinfo {year} {2013})}\BibitemShut
  {NoStop}%
\bibitem [{\citenamefont {Poh}\ \emph {et~al.}(2015)\citenamefont {Poh},
  \citenamefont {Joshi}, \citenamefont {Cer{\`e}}, \citenamefont {Cabello},\
  and\ \citenamefont {Kurtsiefer}}]{poh2015approaching}%
  \BibitemOpen
  \bibfield  {author} {\bibinfo {author} {\bibfnamefont {H.~S.}\ \bibnamefont
  {Poh}}, \bibinfo {author} {\bibfnamefont {S.~K.}\ \bibnamefont {Joshi}},
  \bibinfo {author} {\bibfnamefont {A.}~\bibnamefont {Cer{\`e}}}, \bibinfo
  {author} {\bibfnamefont {A.}~\bibnamefont {Cabello}}, \ and\ \bibinfo
  {author} {\bibfnamefont {C.}~\bibnamefont {Kurtsiefer}},\ }\href@noop {}
  {\bibfield  {journal} {\bibinfo  {journal} {Physical review letters}\
  }\textbf {\bibinfo {volume} {115}},\ \bibinfo {pages} {180408} (\bibinfo
  {year} {2015})}\BibitemShut {NoStop}%
\bibitem [{\citenamefont {Kwiat}\ \emph {et~al.}(1995)\citenamefont {Kwiat},
  \citenamefont {Mattle}, \citenamefont {Weinfurter}, \citenamefont
  {Zeilinger}, \citenamefont {Sergienko},\ and\ \citenamefont
  {Shih}}]{kwiat1995new}%
  \BibitemOpen
  \bibfield  {author} {\bibinfo {author} {\bibfnamefont {P.~G.}\ \bibnamefont
  {Kwiat}}, \bibinfo {author} {\bibfnamefont {K.}~\bibnamefont {Mattle}},
  \bibinfo {author} {\bibfnamefont {H.}~\bibnamefont {Weinfurter}}, \bibinfo
  {author} {\bibfnamefont {A.}~\bibnamefont {Zeilinger}}, \bibinfo {author}
  {\bibfnamefont {A.~V.}\ \bibnamefont {Sergienko}}, \ and\ \bibinfo {author}
  {\bibfnamefont {Y.}~\bibnamefont {Shih}},\ }\href@noop {} {\bibfield
  {journal} {\bibinfo  {journal} {Physical Review Letters}\ }\textbf {\bibinfo
  {volume} {75}},\ \bibinfo {pages} {4337} (\bibinfo {year}
  {1995})}\BibitemShut {NoStop}%
\bibitem [{\citenamefont {Kwiat}\ \emph {et~al.}(1999)\citenamefont {Kwiat},
  \citenamefont {Waks}, \citenamefont {White}, \citenamefont {Appelbaum},\ and\
  \citenamefont {Eberhard}}]{kwiat1999ultrabright}%
  \BibitemOpen
  \bibfield  {author} {\bibinfo {author} {\bibfnamefont {P.~G.}\ \bibnamefont
  {Kwiat}}, \bibinfo {author} {\bibfnamefont {E.}~\bibnamefont {Waks}},
  \bibinfo {author} {\bibfnamefont {A.~G.}\ \bibnamefont {White}}, \bibinfo
  {author} {\bibfnamefont {I.}~\bibnamefont {Appelbaum}}, \ and\ \bibinfo
  {author} {\bibfnamefont {P.~H.}\ \bibnamefont {Eberhard}},\ }\href@noop {}
  {\bibfield  {journal} {\bibinfo  {journal} {Physical Review A}\ }\textbf
  {\bibinfo {volume} {60}},\ \bibinfo {pages} {R773} (\bibinfo {year}
  {1999})}\BibitemShut {NoStop}%
\bibitem [{\citenamefont {Shi}\ and\ \citenamefont
  {Tomita}(2004)}]{shi2004generation}%
  \BibitemOpen
  \bibfield  {author} {\bibinfo {author} {\bibfnamefont {B.-S.}\ \bibnamefont
  {Shi}}\ and\ \bibinfo {author} {\bibfnamefont {A.}~\bibnamefont {Tomita}},\
  }\href@noop {} {\bibfield  {journal} {\bibinfo  {journal} {Physical Review
  A}\ }\textbf {\bibinfo {volume} {69}},\ \bibinfo {pages} {013803} (\bibinfo
  {year} {2004})}\BibitemShut {NoStop}%
\bibitem [{\citenamefont {Fiorentino}\ \emph {et~al.}(2005)\citenamefont
  {Fiorentino}, \citenamefont {Kuklewicz},\ and\ \citenamefont
  {Wong}}]{fiorentino2005source}%
  \BibitemOpen
  \bibfield  {author} {\bibinfo {author} {\bibfnamefont {M.}~\bibnamefont
  {Fiorentino}}, \bibinfo {author} {\bibfnamefont {C.~E.}\ \bibnamefont
  {Kuklewicz}}, \ and\ \bibinfo {author} {\bibfnamefont {F.~N.}\ \bibnamefont
  {Wong}},\ }\href@noop {} {\bibfield  {journal} {\bibinfo  {journal} {Optics
  Express}\ }\textbf {\bibinfo {volume} {13}},\ \bibinfo {pages} {127}
  (\bibinfo {year} {2005})}\BibitemShut {NoStop}%
\bibitem [{\citenamefont {Kim}\ \emph {et~al.}(2006)\citenamefont {Kim},
  \citenamefont {Fiorentino},\ and\ \citenamefont {Wong}}]{kim2006phase}%
  \BibitemOpen
  \bibfield  {author} {\bibinfo {author} {\bibfnamefont {T.}~\bibnamefont
  {Kim}}, \bibinfo {author} {\bibfnamefont {M.}~\bibnamefont {Fiorentino}}, \
  and\ \bibinfo {author} {\bibfnamefont {F.~N.~C.}\ \bibnamefont {Wong}},\
  }\href@noop {} {\bibfield  {journal} {\bibinfo  {journal} {Physical Review
  A}\ }\textbf {\bibinfo {volume} {73}},\ \bibinfo {pages} {012316} (\bibinfo
  {year} {2006})}\BibitemShut {NoStop}%
\bibitem [{\citenamefont {Lee}\ \emph {et~al.}(2016)\citenamefont {Lee},
  \citenamefont {Kim}, \citenamefont {Cha},\ and\ \citenamefont
  {Moon}}]{lee2016polarization}%
  \BibitemOpen
  \bibfield  {author} {\bibinfo {author} {\bibfnamefont {S.~M.}\ \bibnamefont
  {Lee}}, \bibinfo {author} {\bibfnamefont {H.}~\bibnamefont {Kim}}, \bibinfo
  {author} {\bibfnamefont {M.}~\bibnamefont {Cha}}, \ and\ \bibinfo {author}
  {\bibfnamefont {H.~S.}\ \bibnamefont {Moon}},\ }\href@noop {} {\bibfield
  {journal} {\bibinfo  {journal} {Optics express}\ }\textbf {\bibinfo {volume}
  {24}},\ \bibinfo {pages} {2941} (\bibinfo {year} {2016})}\BibitemShut
  {NoStop}%
\bibitem [{\citenamefont {Villar}\ \emph {et~al.}(2018)\citenamefont {Villar},
  \citenamefont {Lohrmann},\ and\ \citenamefont
  {Ling}}]{villar2018experimental}%
  \BibitemOpen
  \bibfield  {author} {\bibinfo {author} {\bibfnamefont {A.}~\bibnamefont
  {Villar}}, \bibinfo {author} {\bibfnamefont {A.}~\bibnamefont {Lohrmann}}, \
  and\ \bibinfo {author} {\bibfnamefont {A.}~\bibnamefont {Ling}},\ }\href@noop
  {} {\bibfield  {journal} {\bibinfo  {journal} {Optics express}\ }\textbf
  {\bibinfo {volume} {26}},\ \bibinfo {pages} {12396} (\bibinfo {year}
  {2018})}\BibitemShut {NoStop}%
\bibitem [{\citenamefont {Fedrizzi}\ \emph {et~al.}(2007)\citenamefont
  {Fedrizzi}, \citenamefont {Herbst}, \citenamefont {Poppe}, \citenamefont
  {Jennewein},\ and\ \citenamefont {Zeilinger}}]{fedrizzi2007wavelength}%
  \BibitemOpen
  \bibfield  {author} {\bibinfo {author} {\bibfnamefont {A.}~\bibnamefont
  {Fedrizzi}}, \bibinfo {author} {\bibfnamefont {T.}~\bibnamefont {Herbst}},
  \bibinfo {author} {\bibfnamefont {A.}~\bibnamefont {Poppe}}, \bibinfo
  {author} {\bibfnamefont {T.}~\bibnamefont {Jennewein}}, \ and\ \bibinfo
  {author} {\bibfnamefont {A.}~\bibnamefont {Zeilinger}},\ }\href@noop {}
  {\bibfield  {journal} {\bibinfo  {journal} {Optics Express}\ }\textbf
  {\bibinfo {volume} {15}},\ \bibinfo {pages} {15377} (\bibinfo {year}
  {2007})}\BibitemShut {NoStop}%
\bibitem [{\citenamefont {Steinlechner}\ \emph {et~al.}(2012)\citenamefont
  {Steinlechner}, \citenamefont {Trojek}, \citenamefont {Jofre}, \citenamefont
  {Weier}, \citenamefont {Perez}, \citenamefont {Jennewein}, \citenamefont
  {Ursin}, \citenamefont {Rarity}, \citenamefont {Mitchell}, \citenamefont
  {Torres} \emph {et~al.}}]{steinlechner2012high}%
  \BibitemOpen
  \bibfield  {author} {\bibinfo {author} {\bibfnamefont {F.}~\bibnamefont
  {Steinlechner}}, \bibinfo {author} {\bibfnamefont {P.}~\bibnamefont
  {Trojek}}, \bibinfo {author} {\bibfnamefont {M.}~\bibnamefont {Jofre}},
  \bibinfo {author} {\bibfnamefont {H.}~\bibnamefont {Weier}}, \bibinfo
  {author} {\bibfnamefont {D.}~\bibnamefont {Perez}}, \bibinfo {author}
  {\bibfnamefont {T.}~\bibnamefont {Jennewein}}, \bibinfo {author}
  {\bibfnamefont {R.}~\bibnamefont {Ursin}}, \bibinfo {author} {\bibfnamefont
  {J.}~\bibnamefont {Rarity}}, \bibinfo {author} {\bibfnamefont {M.~W.}\
  \bibnamefont {Mitchell}}, \bibinfo {author} {\bibfnamefont {J.~P.}\
  \bibnamefont {Torres}},  \emph {et~al.},\ }\href@noop {} {\bibfield
  {journal} {\bibinfo  {journal} {Optics express}\ }\textbf {\bibinfo {volume}
  {20}},\ \bibinfo {pages} {9640} (\bibinfo {year} {2012})}\BibitemShut
  {NoStop}%
\bibitem [{\citenamefont {Steinlechner}\ \emph {et~al.}(2014)\citenamefont
  {Steinlechner}, \citenamefont {Gilaberte}, \citenamefont {Jofre},
  \citenamefont {Scheidl}, \citenamefont {Torres}, \citenamefont {Pruneri},\
  and\ \citenamefont {Ursin}}]{steinlechner2014efficient}%
  \BibitemOpen
  \bibfield  {author} {\bibinfo {author} {\bibfnamefont {F.}~\bibnamefont
  {Steinlechner}}, \bibinfo {author} {\bibfnamefont {M.}~\bibnamefont
  {Gilaberte}}, \bibinfo {author} {\bibfnamefont {M.}~\bibnamefont {Jofre}},
  \bibinfo {author} {\bibfnamefont {T.}~\bibnamefont {Scheidl}}, \bibinfo
  {author} {\bibfnamefont {J.~P.}\ \bibnamefont {Torres}}, \bibinfo {author}
  {\bibfnamefont {V.}~\bibnamefont {Pruneri}}, \ and\ \bibinfo {author}
  {\bibfnamefont {R.}~\bibnamefont {Ursin}},\ }\href@noop {} {\bibfield
  {journal} {\bibinfo  {journal} {JOSA B}\ }\textbf {\bibinfo {volume} {31}},\
  \bibinfo {pages} {2068} (\bibinfo {year} {2014})}\BibitemShut {NoStop}%
\bibitem [{\citenamefont {Jabir}\ and\ \citenamefont
  {Samanta}(2017)}]{jabir2017robust}%
  \BibitemOpen
  \bibfield  {author} {\bibinfo {author} {\bibfnamefont {M.}~\bibnamefont
  {Jabir}}\ and\ \bibinfo {author} {\bibfnamefont {G.}~\bibnamefont
  {Samanta}},\ }\href@noop {} {\bibfield  {journal} {\bibinfo  {journal}
  {Scientific Reports}\ }\textbf {\bibinfo {volume} {7}},\ \bibinfo {pages}
  {12613} (\bibinfo {year} {2017})}\BibitemShut {NoStop}%
\bibitem [{\citenamefont {Dietz}\ \emph {et~al.}(2016)\citenamefont {Dietz},
  \citenamefont {M{\"u}ller}, \citenamefont {Krei{\ss}l}, \citenamefont
  {Herzog}, \citenamefont {Kroh}, \citenamefont {Ahlrichs},\ and\ \citenamefont
  {Benson}}]{dietz2016folded}%
  \BibitemOpen
  \bibfield  {author} {\bibinfo {author} {\bibfnamefont {O.}~\bibnamefont
  {Dietz}}, \bibinfo {author} {\bibfnamefont {C.}~\bibnamefont {M{\"u}ller}},
  \bibinfo {author} {\bibfnamefont {T.}~\bibnamefont {Krei{\ss}l}}, \bibinfo
  {author} {\bibfnamefont {U.}~\bibnamefont {Herzog}}, \bibinfo {author}
  {\bibfnamefont {T.}~\bibnamefont {Kroh}}, \bibinfo {author} {\bibfnamefont
  {A.}~\bibnamefont {Ahlrichs}}, \ and\ \bibinfo {author} {\bibfnamefont
  {O.}~\bibnamefont {Benson}},\ }\href@noop {} {\bibfield  {journal} {\bibinfo
  {journal} {Applied Physics B}\ }\textbf {\bibinfo {volume} {122}},\ \bibinfo
  {pages} {33} (\bibinfo {year} {2016})}\BibitemShut {NoStop}%
\bibitem [{\citenamefont {Fiorentino}\ \emph {et~al.}(2007)\citenamefont
  {Fiorentino}, \citenamefont {Spillane}, \citenamefont {Beausoleil},
  \citenamefont {Roberts}, \citenamefont {Battle},\ and\ \citenamefont
  {Munro}}]{fiorentino2007spontaneous}%
  \BibitemOpen
  \bibfield  {author} {\bibinfo {author} {\bibfnamefont {M.}~\bibnamefont
  {Fiorentino}}, \bibinfo {author} {\bibfnamefont {S.~M.}\ \bibnamefont
  {Spillane}}, \bibinfo {author} {\bibfnamefont {R.~G.}\ \bibnamefont
  {Beausoleil}}, \bibinfo {author} {\bibfnamefont {T.~D.}\ \bibnamefont
  {Roberts}}, \bibinfo {author} {\bibfnamefont {P.}~\bibnamefont {Battle}}, \
  and\ \bibinfo {author} {\bibfnamefont {M.~W.}\ \bibnamefont {Munro}},\
  }\href@noop {} {\bibfield  {journal} {\bibinfo  {journal} {Optics express}\
  }\textbf {\bibinfo {volume} {15}},\ \bibinfo {pages} {7479} (\bibinfo {year}
  {2007})}\BibitemShut {NoStop}%
\bibitem [{\citenamefont {Sohler}\ \emph {et~al.}(2012)\citenamefont {Sohler},
  \citenamefont {Herrmann}, \citenamefont {Ricken}, \citenamefont {Quiring},
  \citenamefont {George}, \citenamefont {Pal}, \citenamefont {Yang},
  \citenamefont {Luo}, \citenamefont {Silberhorn}, \citenamefont {Kaiser} \emph
  {et~al.}}]{sohler2012integrated}%
  \BibitemOpen
  \bibfield  {author} {\bibinfo {author} {\bibfnamefont {W.}~\bibnamefont
  {Sohler}}, \bibinfo {author} {\bibfnamefont {H.}~\bibnamefont {Herrmann}},
  \bibinfo {author} {\bibfnamefont {R.}~\bibnamefont {Ricken}}, \bibinfo
  {author} {\bibfnamefont {V.}~\bibnamefont {Quiring}}, \bibinfo {author}
  {\bibfnamefont {M.}~\bibnamefont {George}}, \bibinfo {author} {\bibfnamefont
  {S.}~\bibnamefont {Pal}}, \bibinfo {author} {\bibfnamefont {X.}~\bibnamefont
  {Yang}}, \bibinfo {author} {\bibfnamefont {K.~H.}\ \bibnamefont {Luo}},
  \bibinfo {author} {\bibfnamefont {C.}~\bibnamefont {Silberhorn}}, \bibinfo
  {author} {\bibfnamefont {F.}~\bibnamefont {Kaiser}},  \emph {et~al.},\ }in\
  \href@noop {} {\emph {\bibinfo {booktitle} {Information Optoelectronics,
  Nanofabrication and Testing}}}\ (\bibinfo {organization} {Optical Society of
  America},\ \bibinfo {year} {2012})\ pp.\ \bibinfo {pages}
  {IF1A--1}\BibitemShut {NoStop}%
\bibitem [{\citenamefont {Krapick}\ \emph {et~al.}(2013)\citenamefont
  {Krapick}, \citenamefont {Herrmann}, \citenamefont {Quiring}, \citenamefont
  {Brecht}, \citenamefont {Suche},\ and\ \citenamefont
  {Silberhorn}}]{krapick2013efficient}%
  \BibitemOpen
  \bibfield  {author} {\bibinfo {author} {\bibfnamefont {S.}~\bibnamefont
  {Krapick}}, \bibinfo {author} {\bibfnamefont {H.}~\bibnamefont {Herrmann}},
  \bibinfo {author} {\bibfnamefont {V.}~\bibnamefont {Quiring}}, \bibinfo
  {author} {\bibfnamefont {B.}~\bibnamefont {Brecht}}, \bibinfo {author}
  {\bibfnamefont {H.}~\bibnamefont {Suche}}, \ and\ \bibinfo {author}
  {\bibfnamefont {C.}~\bibnamefont {Silberhorn}},\ }\href@noop {} {\bibfield
  {journal} {\bibinfo  {journal} {New Journal of Physics}\ }\textbf {\bibinfo
  {volume} {15}},\ \bibinfo {pages} {033010} (\bibinfo {year}
  {2013})}\BibitemShut {NoStop}%
\bibitem [{\citenamefont {Vergyris}\ \emph {et~al.}(2017)\citenamefont
  {Vergyris}, \citenamefont {Kaiser}, \citenamefont {Gouzien}, \citenamefont
  {Sauder}, \citenamefont {Lunghi},\ and\ \citenamefont
  {Tanzilli}}]{vergyris2017fully}%
  \BibitemOpen
  \bibfield  {author} {\bibinfo {author} {\bibfnamefont {P.}~\bibnamefont
  {Vergyris}}, \bibinfo {author} {\bibfnamefont {F.}~\bibnamefont {Kaiser}},
  \bibinfo {author} {\bibfnamefont {E.}~\bibnamefont {Gouzien}}, \bibinfo
  {author} {\bibfnamefont {G.}~\bibnamefont {Sauder}}, \bibinfo {author}
  {\bibfnamefont {T.}~\bibnamefont {Lunghi}}, \ and\ \bibinfo {author}
  {\bibfnamefont {S.}~\bibnamefont {Tanzilli}},\ }\href@noop {} {\bibfield
  {journal} {\bibinfo  {journal} {Quantum Science and Technology}\ }\textbf
  {\bibinfo {volume} {2}},\ \bibinfo {pages} {024007} (\bibinfo {year}
  {2017})}\BibitemShut {NoStop}%
\bibitem [{\citenamefont {Matsuda}\ \emph {et~al.}(2012)\citenamefont
  {Matsuda}, \citenamefont {Le~Jeannic}, \citenamefont {Fukuda}, \citenamefont
  {Tsuchizawa}, \citenamefont {Munro}, \citenamefont {Shimizu}, \citenamefont
  {Yamada}, \citenamefont {Tokura},\ and\ \citenamefont
  {Takesue}}]{matsuda2012monolithically}%
  \BibitemOpen
  \bibfield  {author} {\bibinfo {author} {\bibfnamefont {N.}~\bibnamefont
  {Matsuda}}, \bibinfo {author} {\bibfnamefont {H.}~\bibnamefont {Le~Jeannic}},
  \bibinfo {author} {\bibfnamefont {H.}~\bibnamefont {Fukuda}}, \bibinfo
  {author} {\bibfnamefont {T.}~\bibnamefont {Tsuchizawa}}, \bibinfo {author}
  {\bibfnamefont {W.~J.}\ \bibnamefont {Munro}}, \bibinfo {author}
  {\bibfnamefont {K.}~\bibnamefont {Shimizu}}, \bibinfo {author} {\bibfnamefont
  {K.}~\bibnamefont {Yamada}}, \bibinfo {author} {\bibfnamefont
  {Y.}~\bibnamefont {Tokura}}, \ and\ \bibinfo {author} {\bibfnamefont
  {H.}~\bibnamefont {Takesue}},\ }\href@noop {} {\bibfield  {journal} {\bibinfo
   {journal} {Scientific reports}\ }\textbf {\bibinfo {volume} {2}},\ \bibinfo
  {pages} {817} (\bibinfo {year} {2012})}\BibitemShut {NoStop}%
\bibitem [{\citenamefont {Clausen}\ \emph {et~al.}(2014)\citenamefont
  {Clausen}, \citenamefont {Bussieres}, \citenamefont {Tiranov}, \citenamefont
  {Herrmann}, \citenamefont {Silberhorn}, \citenamefont {Sohler}, \citenamefont
  {Afzelius},\ and\ \citenamefont {Gisin}}]{clausen2014source}%
  \BibitemOpen
  \bibfield  {author} {\bibinfo {author} {\bibfnamefont {C.}~\bibnamefont
  {Clausen}}, \bibinfo {author} {\bibfnamefont {F.}~\bibnamefont {Bussieres}},
  \bibinfo {author} {\bibfnamefont {A.}~\bibnamefont {Tiranov}}, \bibinfo
  {author} {\bibfnamefont {H.}~\bibnamefont {Herrmann}}, \bibinfo {author}
  {\bibfnamefont {C.}~\bibnamefont {Silberhorn}}, \bibinfo {author}
  {\bibfnamefont {W.}~\bibnamefont {Sohler}}, \bibinfo {author} {\bibfnamefont
  {M.}~\bibnamefont {Afzelius}}, \ and\ \bibinfo {author} {\bibfnamefont
  {N.}~\bibnamefont {Gisin}},\ }\href@noop {} {\bibfield  {journal} {\bibinfo
  {journal} {New Journal of Physics}\ }\textbf {\bibinfo {volume} {16}},\
  \bibinfo {pages} {093058} (\bibinfo {year} {2014})}\BibitemShut {NoStop}%
\bibitem [{\citenamefont {Steinlechner}(2015)}]{steinlechner2015sources}%
  \BibitemOpen
  \bibfield  {author} {\bibinfo {author} {\bibfnamefont {F.}~\bibnamefont
  {Steinlechner}},\ }\href@noop {} {\  (\bibinfo {year} {2015})}\BibitemShut
  {NoStop}%
\bibitem [{\citenamefont {Chen}\ \emph {et~al.}(2007)\citenamefont {Chen},
  \citenamefont {Lee},\ and\ \citenamefont {Kumar}}]{chen2007deterministic}%
  \BibitemOpen
  \bibfield  {author} {\bibinfo {author} {\bibfnamefont {J.}~\bibnamefont
  {Chen}}, \bibinfo {author} {\bibfnamefont {K.~F.}\ \bibnamefont {Lee}}, \
  and\ \bibinfo {author} {\bibfnamefont {P.}~\bibnamefont {Kumar}},\
  }\href@noop {} {\bibfield  {journal} {\bibinfo  {journal} {Physical Review
  A}\ }\textbf {\bibinfo {volume} {76}},\ \bibinfo {pages} {031804} (\bibinfo
  {year} {2007})}\BibitemShut {NoStop}%
\bibitem [{\citenamefont {Jin}\ \emph {et~al.}(2014)\citenamefont {Jin},
  \citenamefont {Liu}, \citenamefont {Xu}, \citenamefont {Xia}, \citenamefont
  {Zhong}, \citenamefont {Yuan}, \citenamefont {Zhou}, \citenamefont {Gong},
  \citenamefont {Wang},\ and\ \citenamefont {Zhu}}]{jin2014chip}%
  \BibitemOpen
  \bibfield  {author} {\bibinfo {author} {\bibfnamefont {H.}~\bibnamefont
  {Jin}}, \bibinfo {author} {\bibfnamefont {F.~M.}\ \bibnamefont {Liu}},
  \bibinfo {author} {\bibfnamefont {P.}~\bibnamefont {Xu}}, \bibinfo {author}
  {\bibfnamefont {J.~L.}\ \bibnamefont {Xia}}, \bibinfo {author} {\bibfnamefont
  {M.~L.}\ \bibnamefont {Zhong}}, \bibinfo {author} {\bibfnamefont
  {Y.}~\bibnamefont {Yuan}}, \bibinfo {author} {\bibfnamefont {J.~W.}\
  \bibnamefont {Zhou}}, \bibinfo {author} {\bibfnamefont {Y.~X.}\ \bibnamefont
  {Gong}}, \bibinfo {author} {\bibfnamefont {W.}~\bibnamefont {Wang}}, \ and\
  \bibinfo {author} {\bibfnamefont {S.~N.}\ \bibnamefont {Zhu}},\ }\href@noop
  {} {\bibfield  {journal} {\bibinfo  {journal} {Physical review letters}\
  }\textbf {\bibinfo {volume} {113}},\ \bibinfo {pages} {103601} (\bibinfo
  {year} {2014})}\BibitemShut {NoStop}%
\bibitem [{\citenamefont {Marchildon}\ and\ \citenamefont
  {Helmy}(2016)}]{marchildon2016deterministic}%
  \BibitemOpen
  \bibfield  {author} {\bibinfo {author} {\bibfnamefont {R.~P.}\ \bibnamefont
  {Marchildon}}\ and\ \bibinfo {author} {\bibfnamefont {A.~S.}\ \bibnamefont
  {Helmy}},\ }\href@noop {} {\bibfield  {journal} {\bibinfo  {journal} {Laser
  \& Photonics Reviews}\ }\textbf {\bibinfo {volume} {10}},\ \bibinfo {pages}
  {245} (\bibinfo {year} {2016})}\BibitemShut {NoStop}%
\bibitem [{\citenamefont {Yin}\ \emph {et~al.}(2017)\citenamefont {Yin},
  \citenamefont {Cao}, \citenamefont {Li}, \citenamefont {Liao}, \citenamefont
  {Zhang}, \citenamefont {Ren}, \citenamefont {Cai}, \citenamefont {Liu},
  \citenamefont {Li}, \citenamefont {Dai} \emph {et~al.}}]{yin2017satellite}%
  \BibitemOpen
  \bibfield  {author} {\bibinfo {author} {\bibfnamefont {J.}~\bibnamefont
  {Yin}}, \bibinfo {author} {\bibfnamefont {Y.}~\bibnamefont {Cao}}, \bibinfo
  {author} {\bibfnamefont {Y.-H.}\ \bibnamefont {Li}}, \bibinfo {author}
  {\bibfnamefont {S.-K.}\ \bibnamefont {Liao}}, \bibinfo {author}
  {\bibfnamefont {L.}~\bibnamefont {Zhang}}, \bibinfo {author} {\bibfnamefont
  {J.-G.}\ \bibnamefont {Ren}}, \bibinfo {author} {\bibfnamefont {W.-Q.}\
  \bibnamefont {Cai}}, \bibinfo {author} {\bibfnamefont {W.-Y.}\ \bibnamefont
  {Liu}}, \bibinfo {author} {\bibfnamefont {B.}~\bibnamefont {Li}}, \bibinfo
  {author} {\bibfnamefont {H.}~\bibnamefont {Dai}},  \emph {et~al.},\
  }\href@noop {} {\bibfield  {journal} {\bibinfo  {journal} {Science}\ }\textbf
  {\bibinfo {volume} {356}},\ \bibinfo {pages} {1140} (\bibinfo {year}
  {2017})}\BibitemShut {NoStop}%
\bibitem [{\citenamefont {Liao}\ \emph
  {et~al.}(2017{\natexlab{a}})\citenamefont {Liao}, \citenamefont {Cai},
  \citenamefont {Liu}, \citenamefont {Zhang}, \citenamefont {Li}, \citenamefont
  {Ren}, \citenamefont {Yin}, \citenamefont {Shen}, \citenamefont {Cao},
  \citenamefont {Li} \emph {et~al.}}]{liao2017satellite}%
  \BibitemOpen
  \bibfield  {author} {\bibinfo {author} {\bibfnamefont {S.-K.}\ \bibnamefont
  {Liao}}, \bibinfo {author} {\bibfnamefont {W.-Q.}\ \bibnamefont {Cai}},
  \bibinfo {author} {\bibfnamefont {W.-Y.}\ \bibnamefont {Liu}}, \bibinfo
  {author} {\bibfnamefont {L.}~\bibnamefont {Zhang}}, \bibinfo {author}
  {\bibfnamefont {Y.}~\bibnamefont {Li}}, \bibinfo {author} {\bibfnamefont
  {J.-G.}\ \bibnamefont {Ren}}, \bibinfo {author} {\bibfnamefont
  {J.}~\bibnamefont {Yin}}, \bibinfo {author} {\bibfnamefont {Q.}~\bibnamefont
  {Shen}}, \bibinfo {author} {\bibfnamefont {Y.}~\bibnamefont {Cao}}, \bibinfo
  {author} {\bibfnamefont {Z.-P.}\ \bibnamefont {Li}},  \emph {et~al.},\
  }\href@noop {} {\bibfield  {journal} {\bibinfo  {journal} {Nature}\ }\textbf
  {\bibinfo {volume} {549}},\ \bibinfo {pages} {43} (\bibinfo {year}
  {2017}{\natexlab{a}})}\BibitemShut {NoStop}%
\bibitem [{\citenamefont {Liao}\ \emph
  {et~al.}(2017{\natexlab{b}})\citenamefont {Liao}, \citenamefont {Yong},
  \citenamefont {Liu}, \citenamefont {Shentu}, \citenamefont {Li},
  \citenamefont {Lin}, \citenamefont {Dai}, \citenamefont {Zhao}, \citenamefont
  {Li}, \citenamefont {Guan} \emph {et~al.}}]{liao2017long}%
  \BibitemOpen
  \bibfield  {author} {\bibinfo {author} {\bibfnamefont {S.-K.}\ \bibnamefont
  {Liao}}, \bibinfo {author} {\bibfnamefont {H.-L.}\ \bibnamefont {Yong}},
  \bibinfo {author} {\bibfnamefont {C.}~\bibnamefont {Liu}}, \bibinfo {author}
  {\bibfnamefont {G.-L.}\ \bibnamefont {Shentu}}, \bibinfo {author}
  {\bibfnamefont {D.-D.}\ \bibnamefont {Li}}, \bibinfo {author} {\bibfnamefont
  {J.}~\bibnamefont {Lin}}, \bibinfo {author} {\bibfnamefont {H.}~\bibnamefont
  {Dai}}, \bibinfo {author} {\bibfnamefont {S.-Q.}\ \bibnamefont {Zhao}},
  \bibinfo {author} {\bibfnamefont {B.}~\bibnamefont {Li}}, \bibinfo {author}
  {\bibfnamefont {J.-Y.}\ \bibnamefont {Guan}},  \emph {et~al.},\ }\href@noop
  {} {\bibfield  {journal} {\bibinfo  {journal} {Nature Photonics}\ }\textbf
  {\bibinfo {volume} {11}},\ \bibinfo {pages} {509} (\bibinfo {year}
  {2017}{\natexlab{b}})}\BibitemShut {NoStop}%
\bibitem [{\citenamefont {Scheidl}\ \emph {et~al.}(2013)\citenamefont
  {Scheidl}, \citenamefont {Wille},\ and\ \citenamefont
  {Ursin}}]{scheidl2013quantum}%
  \BibitemOpen
  \bibfield  {author} {\bibinfo {author} {\bibfnamefont {T.}~\bibnamefont
  {Scheidl}}, \bibinfo {author} {\bibfnamefont {E.}~\bibnamefont {Wille}}, \
  and\ \bibinfo {author} {\bibfnamefont {R.}~\bibnamefont {Ursin}},\
  }\href@noop {} {\bibfield  {journal} {\bibinfo  {journal} {New Journal of
  Physics}\ }\textbf {\bibinfo {volume} {15}},\ \bibinfo {pages} {043008}
  (\bibinfo {year} {2013})}\BibitemShut {NoStop}%
\bibitem [{\citenamefont {Wang}\ \emph {et~al.}(2014)\citenamefont {Wang},
  \citenamefont {Guo}, \citenamefont {Zhang},\ and\ \citenamefont
  {Liu}}]{wang2014link}%
  \BibitemOpen
  \bibfield  {author} {\bibinfo {author} {\bibfnamefont {X.}~\bibnamefont
  {Wang}}, \bibinfo {author} {\bibfnamefont {L.}~\bibnamefont {Guo}}, \bibinfo
  {author} {\bibfnamefont {L.}~\bibnamefont {Zhang}}, \ and\ \bibinfo {author}
  {\bibfnamefont {Y.}~\bibnamefont {Liu}},\ }\href@noop {} {\bibfield
  {journal} {\bibinfo  {journal} {Optics Communications}\ }\textbf {\bibinfo
  {volume} {310}},\ \bibinfo {pages} {12} (\bibinfo {year} {2014})}\BibitemShut
  {NoStop}%
\bibitem [{\citenamefont {{Neiman, Yasha}}(2018)}]{neiman2018towards}%
  \BibitemOpen
  \bibfield  {author} {\bibinfo {author} {\bibnamefont {{Neiman, Yasha}}},\
  }\href {\doibase 10.1051/epjconf/201816801007} {\bibfield  {journal}
  {\bibinfo  {journal} {EPJ Web Conf.}\ }\textbf {\bibinfo {volume} {168}},\
  \bibinfo {pages} {01007} (\bibinfo {year} {2018})}\BibitemShut {NoStop}%
\bibitem [{\citenamefont {Ray}\ and\ \citenamefont {van
  Enk}(2011)}]{ray2011verifying}%
  \BibitemOpen
  \bibfield  {author} {\bibinfo {author} {\bibfnamefont {M.~R.}\ \bibnamefont
  {Ray}}\ and\ \bibinfo {author} {\bibfnamefont {S.~J.}\ \bibnamefont {van
  Enk}},\ }\href@noop {} {\bibfield  {journal} {\bibinfo  {journal} {Physical
  Review A}\ }\textbf {\bibinfo {volume} {83}},\ \bibinfo {pages} {042318}
  (\bibinfo {year} {2011})}\BibitemShut {NoStop}%
\bibitem [{\citenamefont {B{\"u}se}\ \emph {et~al.}(2015)\citenamefont
  {B{\"u}se}, \citenamefont {Tischler}, \citenamefont {Juan},\ and\
  \citenamefont {Molina-Terriza}}]{buse2015photons}%
  \BibitemOpen
  \bibfield  {author} {\bibinfo {author} {\bibfnamefont {A.}~\bibnamefont
  {B{\"u}se}}, \bibinfo {author} {\bibfnamefont {N.}~\bibnamefont {Tischler}},
  \bibinfo {author} {\bibfnamefont {M.~L.}\ \bibnamefont {Juan}}, \ and\
  \bibinfo {author} {\bibfnamefont {G.}~\bibnamefont {Molina-Terriza}},\
  }\href@noop {} {\bibfield  {journal} {\bibinfo  {journal} {Journal of
  Optics}\ }\textbf {\bibinfo {volume} {17}},\ \bibinfo {pages} {065201}
  (\bibinfo {year} {2015})}\BibitemShut {NoStop}%
\bibitem [{\citenamefont {Trojek}\ and\ \citenamefont
  {Weinfurter}(2008)}]{trojek2008collinear}%
  \BibitemOpen
  \bibfield  {author} {\bibinfo {author} {\bibfnamefont {P.}~\bibnamefont
  {Trojek}}\ and\ \bibinfo {author} {\bibfnamefont {H.}~\bibnamefont
  {Weinfurter}},\ }\href@noop {} {\bibfield  {journal} {\bibinfo  {journal}
  {Applied Physics Letters}\ }\textbf {\bibinfo {volume} {92}},\ \bibinfo
  {pages} {211103} (\bibinfo {year} {2008})}\BibitemShut {NoStop}%
\bibitem [{\citenamefont {Pan}\ \emph {et~al.}(2003)\citenamefont {Pan},
  \citenamefont {Gasparoni}, \citenamefont {Ursin}, \citenamefont {Weihs},\
  and\ \citenamefont {Zeilinger}}]{pan2003experimental}%
  \BibitemOpen
  \bibfield  {author} {\bibinfo {author} {\bibfnamefont {J.-W.}\ \bibnamefont
  {Pan}}, \bibinfo {author} {\bibfnamefont {S.}~\bibnamefont {Gasparoni}},
  \bibinfo {author} {\bibfnamefont {R.}~\bibnamefont {Ursin}}, \bibinfo
  {author} {\bibfnamefont {G.}~\bibnamefont {Weihs}}, \ and\ \bibinfo {author}
  {\bibfnamefont {A.}~\bibnamefont {Zeilinger}},\ }\href@noop {} {\bibfield
  {journal} {\bibinfo  {journal} {Nature}\ }\textbf {\bibinfo {volume} {423}},\
  \bibinfo {pages} {417} (\bibinfo {year} {2003})}\BibitemShut {NoStop}%
\bibitem [{\citenamefont {Steinlechner}\ \emph {et~al.}(2017)\citenamefont
  {Steinlechner}, \citenamefont {Ecker}, \citenamefont {Fink}, \citenamefont
  {Liu}, \citenamefont {Bavaresco}, \citenamefont {Huber}, \citenamefont
  {Scheidl},\ and\ \citenamefont {Ursin}}]{steinlechner2017distribution}%
  \BibitemOpen
  \bibfield  {author} {\bibinfo {author} {\bibfnamefont {F.}~\bibnamefont
  {Steinlechner}}, \bibinfo {author} {\bibfnamefont {S.}~\bibnamefont {Ecker}},
  \bibinfo {author} {\bibfnamefont {M.}~\bibnamefont {Fink}}, \bibinfo {author}
  {\bibfnamefont {B.}~\bibnamefont {Liu}}, \bibinfo {author} {\bibfnamefont
  {J.}~\bibnamefont {Bavaresco}}, \bibinfo {author} {\bibfnamefont
  {M.}~\bibnamefont {Huber}}, \bibinfo {author} {\bibfnamefont
  {T.}~\bibnamefont {Scheidl}}, \ and\ \bibinfo {author} {\bibfnamefont
  {R.}~\bibnamefont {Ursin}},\ }\href@noop {} {\bibfield  {journal} {\bibinfo
  {journal} {Nature communications}\ }\textbf {\bibinfo {volume} {8}},\
  \bibinfo {pages} {15971} (\bibinfo {year} {2017})}\BibitemShut {NoStop}%
\bibitem [{\citenamefont {Zhang}\ \emph {et~al.}(2016)\citenamefont {Zhang},
  \citenamefont {Roux}, \citenamefont {Konrad}, \citenamefont {Agnew},
  \citenamefont {Leach},\ and\ \citenamefont {Forbes}}]{zhang2016engineering}%
  \BibitemOpen
  \bibfield  {author} {\bibinfo {author} {\bibfnamefont {Y.}~\bibnamefont
  {Zhang}}, \bibinfo {author} {\bibfnamefont {F.~S.}\ \bibnamefont {Roux}},
  \bibinfo {author} {\bibfnamefont {T.}~\bibnamefont {Konrad}}, \bibinfo
  {author} {\bibfnamefont {M.}~\bibnamefont {Agnew}}, \bibinfo {author}
  {\bibfnamefont {J.}~\bibnamefont {Leach}}, \ and\ \bibinfo {author}
  {\bibfnamefont {A.}~\bibnamefont {Forbes}},\ }\href@noop {} {\bibfield
  {journal} {\bibinfo  {journal} {Science advances}\ }\textbf {\bibinfo
  {volume} {2}},\ \bibinfo {pages} {e1501165} (\bibinfo {year}
  {2016})}\BibitemShut {NoStop}%
\bibitem [{\citenamefont {Krenn}\ \emph {et~al.}(2017)\citenamefont {Krenn},
  \citenamefont {Hochrainer}, \citenamefont {Lahiri},\ and\ \citenamefont
  {Zeilinger}}]{krenn2017entanglement}%
  \BibitemOpen
  \bibfield  {author} {\bibinfo {author} {\bibfnamefont {M.}~\bibnamefont
  {Krenn}}, \bibinfo {author} {\bibfnamefont {A.}~\bibnamefont {Hochrainer}},
  \bibinfo {author} {\bibfnamefont {M.}~\bibnamefont {Lahiri}}, \ and\ \bibinfo
  {author} {\bibfnamefont {A.}~\bibnamefont {Zeilinger}},\ }\href@noop {}
  {\bibfield  {journal} {\bibinfo  {journal} {Physical review letters}\
  }\textbf {\bibinfo {volume} {118}},\ \bibinfo {pages} {080401} (\bibinfo
  {year} {2017})}\BibitemShut {NoStop}%
\bibitem [{\citenamefont {Scheidl}\ \emph {et~al.}(2014)\citenamefont
  {Scheidl}, \citenamefont {Tiefenbacher}, \citenamefont {Prevedel},
  \citenamefont {Steinlechner}, \citenamefont {Ursin},\ and\ \citenamefont
  {Zeilinger}}]{scheidl2014crossed}%
  \BibitemOpen
  \bibfield  {author} {\bibinfo {author} {\bibfnamefont {T.}~\bibnamefont
  {Scheidl}}, \bibinfo {author} {\bibfnamefont {F.}~\bibnamefont
  {Tiefenbacher}}, \bibinfo {author} {\bibfnamefont {R.}~\bibnamefont
  {Prevedel}}, \bibinfo {author} {\bibfnamefont {F.}~\bibnamefont
  {Steinlechner}}, \bibinfo {author} {\bibfnamefont {R.}~\bibnamefont {Ursin}},
  \ and\ \bibinfo {author} {\bibfnamefont {A.}~\bibnamefont {Zeilinger}},\
  }\href@noop {} {\bibfield  {journal} {\bibinfo  {journal} {Physical Review
  A}\ }\textbf {\bibinfo {volume} {89}},\ \bibinfo {pages} {042324} (\bibinfo
  {year} {2014})}\BibitemShut {NoStop}%
\bibitem [{\citenamefont {Sanaka}\ \emph {et~al.}(2006)\citenamefont {Sanaka},
  \citenamefont {Resch},\ and\ \citenamefont
  {Zeilinger}}]{sanaka2006filtering}%
  \BibitemOpen
  \bibfield  {author} {\bibinfo {author} {\bibfnamefont {K.}~\bibnamefont
  {Sanaka}}, \bibinfo {author} {\bibfnamefont {K.~J.}\ \bibnamefont {Resch}}, \
  and\ \bibinfo {author} {\bibfnamefont {A.}~\bibnamefont {Zeilinger}},\
  }\href@noop {} {\bibfield  {journal} {\bibinfo  {journal} {Physical review
  letters}\ }\textbf {\bibinfo {volume} {96}},\ \bibinfo {pages} {083601}
  (\bibinfo {year} {2006})}\BibitemShut {NoStop}%
\bibitem [{\citenamefont {Resch}\ \emph {et~al.}(2007)\citenamefont {Resch},
  \citenamefont {O’Brien}, \citenamefont {Weinhold}, \citenamefont {Sanaka},
  \citenamefont {Lanyon}, \citenamefont {Langford},\ and\ \citenamefont
  {White}}]{resch2007entanglement}%
  \BibitemOpen
  \bibfield  {author} {\bibinfo {author} {\bibfnamefont {K.~J.}\ \bibnamefont
  {Resch}}, \bibinfo {author} {\bibfnamefont {J.~L.}\ \bibnamefont
  {O’Brien}}, \bibinfo {author} {\bibfnamefont {T.~J.}\ \bibnamefont
  {Weinhold}}, \bibinfo {author} {\bibfnamefont {K.}~\bibnamefont {Sanaka}},
  \bibinfo {author} {\bibfnamefont {B.~P.}\ \bibnamefont {Lanyon}}, \bibinfo
  {author} {\bibfnamefont {N.~K.}\ \bibnamefont {Langford}}, \ and\ \bibinfo
  {author} {\bibfnamefont {A.~G.}\ \bibnamefont {White}},\ }\href@noop {}
  {\bibfield  {journal} {\bibinfo  {journal} {Physical review letters}\
  }\textbf {\bibinfo {volume} {98}},\ \bibinfo {pages} {203602} (\bibinfo
  {year} {2007})}\BibitemShut {NoStop}%
\end{thebibliography}%

\end{document}